\documentclass[
reprint, aps, prb
]{revtex4-2}

\usepackage{graphicx}
\usepackage{dcolumn}
\usepackage{bm}

\begin{document}
\title{Multicellular control of gene networks}

\author{Kyle R. Allison}
\email{kyle.r.allison@emory.edu}
 \affiliation{Department of Medicine, Division of Infectious Diseases, Emory University School of Medicine, Atlanta, GA}

\begin{abstract}
Biological organisms are simple at heart: cells, their basic units, perform a variety of behaviors by expressing proteins from DNA-encoded genes. Gene expression though depends on sets of often-convoluted regulatory interactions known as gene networks, contributing to biology's apparent complexity. Even in \textit{Escherichia coli} K-12, the pioneering model organism of molecular biology, gene networks are complicated and inconsistent with gene expression data. Recent discoveries of multicellular self-organization in \textit{E. coli} suggest a new model for gene regulation that may help in many cases: control of gene networks by multicellular interaction networks, \textit{i.e.} the changing physical and chemical interactions between cells in communities. \textit{E. coli}'s observed dynamics indicate multicellular interactions serve as inputs for key gene networks, thereby producing robust expression of otherwise noisy genes. In turn, multicellular interaction networks are dynamically generated as outputs of gene networks during self-organization. Thus, multicellular self-organization enables uniquely-biological control mechanisms that are not available to individual cells. As further suggested by \textit{E. coli} self-organization, the gene-network outputs from one multicellular stage can be linked, \textit{via} multicellular interaction networks, to serve as the gene-network inputs for later stages, thereby creating \textit{multicellular daisy chains}. Daisy chains are a general model for reliably propagating multicellular communities and, therefore, for predictably determining the gene-network inputs of each cell throughout multicellular self-organization and development. Considerations for several fields are briefly discussed and suggest multicellular daisy chains are a general mechanism for biological organisms to control cell behavior and to be simpler than the sum of their parts.
\end{abstract}

\maketitle
Studying lactose metabolism in \textit{E. coli}, Francois Jacob and Jacques Monod illustrated that genes can be turned on and off and their expression regulated by molecular interactions with DNA \cite{RN1} (\textbf{Fig. 1}). The importance was immediately apparent and, along with Andre Lwoff who had uncovered the basics of bacteriophage lysogeny, they received a Nobel Prize shortly after. Only a decade earlier, X-ray diffraction patterns revealed DNA (from \textit{E. coli} and calf) had a helical crystalline structure \cite{RN2, RN3} whose base-pairing chemistry indicated how it is copied \cite{RN4} and hence how genes are inherited by dividing cells. Whereas DNA is quite stable, cell behavior is not and changes often. Part of the promise of gene regulation was that it might elegantly simplify the dynamic behaviors of cells to the crystalline chemistry of their nucleic acids.

Even \textit{E. coli} can be complicated though: a gene could regulate many others or be regulated by many others. This led to "gene regulatory networks," first in theory \cite{RN17} and later in \textit{E. coli} \cite{RN6} (\textbf{Fig. 1}). Such networks indicated biology might need the help of engineers, who by profession discern the dynamics of complicated systems. Useful ideas were proposed, like simulating gene networks on computers \cite{RN7} to model $\lambda$ phage in \textit{E. coli}, whose details had been excellently articulated \cite{RN8}. However, the point of a complicated engineered network is often to ensure simple dynamics: by analogy, the structure of regulatory networks might robustly produce straightforward cell behaviors. This was nicely illustrated for the network governing chemotaxis (cell swimming towards chemicals, see \cite{RN9}), first in theory \cite{RN10} then in \textit{E. coli} \cite{RN11}. The molecular network itself mattered, not just the genes. These and similar developments indicated that documenting, analyzing, and simulating networks could aid in understanding biological organisms, further motivating new systems biology approaches (\textbf{Fig. 1}).

\begin{figure}
\includegraphics{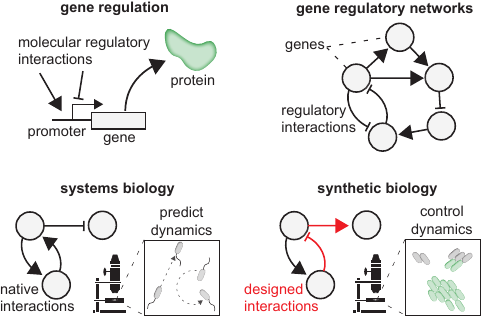}
\caption{Gene regulatory networks and dynamic cell behavior in \textit{E. coli}.}
\end{figure}

Not all systems can be \textit{engineered} however: even small, simple networks can have unpredictable dynamics (see examples in \cite{RN13}). Controlling such systems is often unfeasible. For a system to be \textit{engineer-able}, its components must be arrangeable to produce stable and predictable dynamics. A biological organism would be engineer-able if its gene networks could be rearranged to produce cell behaviors whose dynamics are controllable and predictable (\textbf{Fig. 1}). This was first achieved in \textit{E. coli} with the construction of the genetic toggle switch \cite{RN12}, which together with the repressilator \cite{RN14} launched synthetic biology 25 years ago. Both studies combine regulatory components from the \textit{lac} operon and $\lambda$ phage with green fluorescent protein (\textit{GFP}) \cite{RN18} to design "genetic circuits." Biology being complex, there was no guarantee that synthetic genetic circuits would be predictable. This point was illustrated by the repressilator: it oscillated as designed, but not predictably. The toggle switch proved cell behavior could be engineered by rearranging gene networks, through DNA modification, in line with control theory. This raised the stakes for ongoing efforts to engineer proteins, nucleic acids, metabolites, \textit{etc}. using chemistry and sparked visions of creating new life forms. More concretely, the papers that launched synthetic biology showed that the principles of cell behavior \textit{mirrored} engineering principles, in \textit{E. coli} at least.

Engineered systems tend to repeat variations on common designs; consistently, it was shown that \textit{E. coli}'s complicated transcription factor networks could be refactored as a collection of "network motifs" \cite{RN19}. Such findings suggested that it might be possible to reverse engineer natural gene networks, whether or not their principles fit existing engineering disciplines. Moreover, gene networks might then be inferred by perturbing cells and tracking the resulting expression dynamics, subsequently demonstrated in \textit{E. coli} for antibiotic targets \cite{RN20}. Data from microarrays and sequencing exploded, and gene network inference advanced along with it. Such efforts culminated in a competition and accompanying research to infer gene networks by machine learning \cite{RN21}. Beyond bench-marking gene network inference methods, the resulting "wisdom of crowds" findings showed that the collective insights from a research community outperformed any one group, and nearly half of the predictions tested in \textit{E. coli} were  experimentally validated \cite{RN21}. With the help of machine intelligence, a comprehensive understanding of gene networks appeared to be within grasp.

More than a decade has since passed and our understanding of gene networks in \textit{E. coli} remains incomplete. Despite many advances, limitations remain \cite{RN25}, and substantial discrepancies between \textit{E. coli}’s gene networks and expression have been explicitly pointed out \cite{RN26}. Meanwhile, machine learning inference of gene networks has greatly improved \cite{RN24} and is increasingly applied to decode sophisticated organisms much larger than \textit{E. coli}. I wonder what Jacques Monod would think about all of this, and repeating his oft-repeated statement seems unavoidable: ‘‘Anything that is true of \textit{E. coli} must be true for elephants, except more so’’ \cite{RN27}. The elephant in the room appears to be our limited understanding of how gene networks determine behavior in \textit{E. coli}, to say nothing of higher organisms. Similar reservations have also motivated calls for fresh paradigms for biology \cite{RN44}, like simulating “virtual cells” on computers \cite{RN100, RN101} using modern machine intelligence  \cite{RN102}.

As \textit{E. coli}'s cell behavior can resemble engineering, it might also be useful to revisit reverse engineering \cite{RN55}, \textit{i.e.} uncovering the connections between a system’s inputs and outputs. The outputs of gene networks are gene expression, cell behavior, \textit{etc}. Inputs are often signals from the external environment. For example, when lactose is present, \textit{E. coli} cells express lactose permease (to import it) and $\beta$-galactosidase (to digest it) from the \textit{lac} operon. In this and similar homeostatic networks, outputs functionally correspond to inputs. But not all gene networks are so straightforward. Just as an example, the regulatory architecture of \textit{E. coli}'s \textit{cpx} envelope stress response (a phosphorelay two-component system \cite{RN66}) is more development-like \cite{RN56, RN28}: outputs do not all necessarily correspond to inputs and cells robustly commit to them even if input signals disappear. Ideas like this were extended to argue some gene networks interpret specific inputs as representing an ecological niche and its changes as a whole, first in \textit{E. coli} \cite{RN29} then later more theoretically \cite{RN30}. Such gene networks can only be understood in the context of their inducing environment—which cells perceive as combinations of external inputs. A similar caveat had in fact informed the validation of "wisdom of crowds” predictions in \textit{E. coli} \cite{RN21}: experiments could only be properly designed when a gene network’s inducing conditions were clear. Fortunately, due to extensive prior research, many \textit{E. coli} gene networks do have known inducing conditions and can be reverse engineered by focusing on input signals. This is commonly untrue however for multicellular gene networks, \textit{i.e.} those involved in multicellular communities and biofilms. Somewhere between 10-40 percent of \textit{E. coli}’s genes are differentially expressed in multicellular communities \cite{RN134, RN135}, but their precise inducing conditions during community formation are largely unclear.

From genetic studies (\textit{e.g.} \cite{RN68}), there exists a "parts list" for \textit{E. coli} biofilms, though a model for how these genetic parts worked together was lacking. The related gene networks are critical for colonization and survival, and they are also complicated, with combinatorial inputs and switch-like architecture (\textbf{NB}: summaries of update-to-date knowledge on \textit{E. coli} genes and regulation can be found at EcoCyc \cite{RN23} and RegulonDB \cite{RN22}). Surprisingly, single-cell experiments have often shown that gene expression from many of these networks is stochastic and rare (\textit{e.g.}, \textit{Ag43} \cite{RN31, RN94}, flagella \cite{RN32,RN98}, fimbriae \cite{RN33}, curli \cite{RN34}, and \textit{rpoS} \cite{RN35}). A degree of noise is inherent to gene expression, best demonstrated for \textit{E. coli} in single cells with the \textit{lac} operon \cite{RN133}. From an engineering perspective however, all these noisy gene networks do not add up: why regulate important genes with sophisticated networks just to leave their expression (and cell behavior) to random chance? But gene expression noise can also result from external sources \cite{RN133}, like input signals. As also shown with \textit{E. coli}’s \textit{lac} operon, expression of an inducible gene can be stochastic and rare when cells experience sub-optimal induction \cite{RN38}. Perhaps the noise observed for multicellular genes has resulted from sub-optimal inducing conditions, where the inputs of gene networks are incompletely met. This had largely been unavoidable as suitable methods to \textit{developmentally} track multicellular self-organization in \textit{E. coli} \cite{RN39} were not yet available. Then they were and we found out \textit{E. coli} can self-organize multicellular communities that grow to ~1,000 cells before becoming dormant and attaching to surfaces as clonal units \cite{RN36, RN37}. During multicellular self-organization, in which the actions of individual cells dynamically generate community structure \cite{RN39}, \textit{E. coli}’s otherwise-noisy, multicellular genes reliably perform specific roles in an ordered sequence of multicellular stages (\textbf{Fig. 2}). For some of \textit{E. coli}’s gene networks evidently, \textit{multicellularity} is the inducing condition.

New ideas for microbial multicellularity might be needed to elaborate on this given that \textit{E. coli} is a unicelluar organism, of course, not a multicellular one. As it happens, some such ideas were recently introduced to explore common principles of multicellular self-organization \cite{RN39}. A few are briefly discussed below. \textit{Multicellularity}'s meaning can vary even within evolutionary biology where it is often considered, and standardizing it through evolutionary concepts has been proposed \cite{RN118}. On the other hand, many fields study the mechanisms of multicellular behaviors independent of their evolution. It should then be feasible to have a general theory of multicellularity, or multicellular biology, that applies across multiple disciplines, including evolutionary, developmental, mammalian, plant, bacterial, and synthetic biology \cite{RN39}. Within this framework, \textit{E. coli}’s \textit{multicellularity} is represented as the empirical set of its multicellular behaviors, which to date include self-organization of developmental rosettes, multicellular extension, and multicellular attached dormancy. Comparing behaviors like these between differing organisms has potential to reveal core principles of multicellularity, if they happen to exist, and such approaches have been pioneered in extant protozoa  related to the ancestors of metazoa \cite{RN103, RN105, RN104, RN106, RN139}. Given the diversity in genetics, biochemistry, behaviors \textit{etc}. of biological organisms, further extending such comparisons will require a common basis.

Dynamic graphs can formally capture key geometric properties of communities and illustrate fundamental concepts for multicellular self-organization \cite{RN39}. They are a flexible basis for comparative multicellularity where cells are represented as nodes and their physical and chemical interactions, as edges. Dynamic graphs are characterized by the rules defining how one graph propagates to the next \cite{RN41}. For multicellular graphs (the application of dynamic graphs to multicellular communities) the propagation rules themselves can change dynamically and must be discovered by experimentally observing cells and their changing interactions, rather than extrapolated from the properties of single cells \cite{RN39}. Cells can update and change their community propagation rules by sensing their environment and altering their behavior in response, most obviously through gene regulation. As the position of a cell in a community can determine its behavior and function (generally appreciated for multicellular organisms), predictable cell behavior in a community requires reliable dynamic transitions between its multicellular graphs. In fact, comparing graphs for different community formation behaviors suggests the ability to reliably propagate multicellular graphs is uniquely enabled by clonal multicellular self-organization \cite{RN39}. Dynamic graphs are well-suited to studying multicellular behaviors: they can be easily verified or updated based on empirical observations of communities and, as mathematical constructs, they can also be explored by theoretical development. They have previously helped to characterize multicellular behaviors in yeast \cite{RN130,RN58} and explore topological constraints for communities \cite{RN57, RN123}. Multicellular graphs and the communities they represent are inherently dynamic and changing, but they can approximate a steady state by creating terminal multicellular graphs, wherein cell division is suspended and cell-cell interactions become fixed \cite{RN39}. Such terminal graphs have at times been referred to as "multicellular graphs" without considering the dynamic sequence of multicellular graphs preceding them.

\begin{figure}
\includegraphics{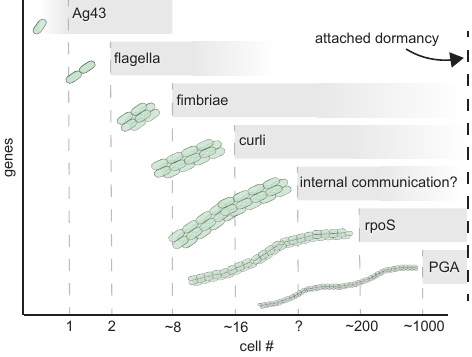}
\caption{Inferred expression dynamics of multicellular genes during \textit{E. coli} self-organization (from data in \cite{RN36, RN37}).}
\end{figure}

Multicellular graphs can be broken down into nodes (cells) and edges (cell-cell interactions). Characterizing the chemical states of cells in multicellular organisms, \textit{e.g.} by single-cell or spatial transcriptomics, can produce critical data, but does not capture the self-organizing behaviors cells perform. These behaviors can only be revealed by observing a community's multicellular dynamics \cite{RN39, RN40}.  Such approaches are used to study cell behavior in animals \cite{RN140, RN141, RN142, RN143} but are rarely applied to bacterial cells, who are commonly expected to lack sophisticated organizational behaviors. In communities, cell behavior and its changes depend on the mechanical and chemical interactions between cells. The combinations of such interactions in multicellular graphs can be thought of as \textit{multicellular interaction networks} (this phrase was also recently applied for tumor micro-environments \cite{RN70}), where the edges between cells are the focus rather than the cells themselves. When sensed, multicellular interaction networks would serve as combinatorial input signals for gene networks. For example, interaction between one cell’s extracellular appendages and another cell’s outer membrane could induce envelope-stress gene networks; metabolite depletion in the inner cavities created within communities could trigger nutrient-stress gene networks; and communication molecules would enable internal coordination \cite{RN39}. \textit{Multicellularity} is then the inducing condition when a gene network produces reliable expression in response to a multicellular interaction network (\textbf{Fig. 3}). This is similar to the idea of positional information for animal developmental biology \cite{RN47} but, instead, is mechanistically framed by gene- and interaction networks and motivated by a bacterium. During multicellular self-organization, the sequential expression of multicellular genes would therefore entail the sequential construction of unique multicellular interaction networks, generated one stage at a time. In \textit{E. coli}, a variety of multicellular interactions have already been revealed (\textbf{Fig. 3}) and the molecular-scale regulation for the related genes has been studied. Putting these pieces together (multicellular interaction networks and gene networks) will illuminate many of \textit{E. coli}’s cell behaviors and will go a long way to completing the picture of gene regulation in this model organism.

While the connections between interaction networks and gene networks mostly remain to be discovered, theoretical insights can be gained by interpreting known gene regulation within the context of self-organization. Some fundamentals first merit mention. In \textit{E. coli}, cell division and gene expression have comparable times scales: cell division takes about 30 minutes, typically enough time for significant \textit{de novo} gene expression. Prior to self-organization, most multicellular genes are repressed in single cells. Expression of some genes could therefore be synchronized with the cell cycle and multicellular interactions could then adapt at each division event of self-organization, (\textit{e.g.}, 2-cell, 4-cell, \textit{etc.}). Rules of multicellular graph propagation could potentially change discretely at each multicellular stage, indicating multicellular graph propagation at times may follow Boolean logic. Expression dynamics also indicate that induction occurs approximately one or more division cycles before a multicellular gene's function is observable. Several genes encode large extracellular appendages that take generations to fully polymerize (\textit{e.g.}, fimbriae, curli, \textit{etc.}), and may be induced multiple division before their function is observable. However, even some of these genes play critical roles before reaching their full functional maturity, exemplified by flagella which only self-organizes rosettes when synthesized \textit{de novo} \cite{RN37}. There are also considerations for how interactions are established between cells. \textit{E. coli} self-organization occurs clonally, meaning all cells descend from a single parental cell, even in 1,000-cell communities \cite{RN36}. Multicellular interactions are often viewed as the result of decisions by separate cells about how to behave. Yet, in clonal communities, many interactions are between sister cells, arising from division of one cell into two. In this case, multicellular interactions are coordinated before division takes place; for example, extracellular surface proteins produced by mother cells would be equally shared by daughter cells, unless a form of asymmetric localization or division was involved. Fluorescent proteins like \textit{GFP} \cite{RN18} are valuable tools here: they allow quantification of gene expression dynamics with minimal disturbance to cells, have been optimized for \textit{E. coli} \cite{RN97}, and were used to create an extensive library of transcriptional reporters for \textit{E. coli} \cite{RN69}. Eventually, there will be a detailed temporal map of gene expression for \textit{E. coli} self-organization that fully connects multicellular genes and the multicellular interactions inducing them. Achieving this will require a considerable amount of multicellular dynamic experiments, but at least the tools and techniques are already available.

\begin{figure}
\includegraphics{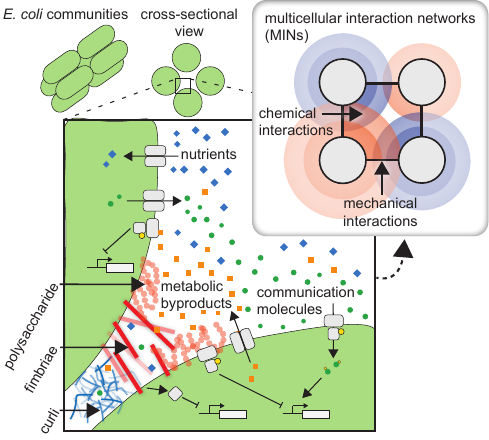}
\caption{A variety of mechanical and chemical interactions exist within multicellular \textit{E. coli} communities (see also \cite{RN39}) and can be represented as multicellular interaction networks (edges) connecting cells (nodes).}
\end{figure}

\textit{E. coli} multicellular self-organization (the stages, behaviors, and genes of which are described in \cite{RN39}) begins when sister cells remain adhered at their poles in environments which allow cell motion but lack sufficient forces to cause cell separation. This adhesion is mediated by the self-recognizing surface protein Antigen 43 (\textit{Ag43}), which binds to identical proteins on neighboring cells \cite{RN146, RN148,RN147, RN149}. The \textit{flu} gene encodes \textit{Ag43} and is negatively regulated by the reduced form of \textit{OxyR}, a key transcriptional regulator responsive to the cell's internal oxidative state \cite{RN150, RN151, RN152}. Under anaerobic conditions, \textit{flu} is fully repressed, possibly explaining the lack of anaerobic biofilm formation by \textit{E. coli} K-12 \cite{RN153}. Conversely, knockout mutation to \textit{oxyR} leads to full induction of \textit{flu}, causing cells to auto-aggregate in static conditions \cite{RN154}. Typically, in aerobic environments, \textit{flu} is expressed at intermediate levels \cite{RN149}, allowing \textit{Ag43}'s presence on cell surfaces to enable the initial step of multicellular self-organization. Additionally, \textit{flu} expression is epigenetically regulated by Dam methylase, which has three binding sites overlapping with the three \textit{OxyR} binding sites in the promoter \cite{RN149}. Binding of \textit{OxyR} prevents Dam methylation, and methylation blocks \textit{OxyR} binding. Hence, \textit{flu} expression depends on competition between constitutively active Dam methylase and reduced \textit{OxyR} during DNA synthesis. The promoter's methylation state locks in expression  until the next round of DNA synthesis, enhancing correlation of \textit{flu} expression in daughter cells.

The \textit{flu} gene network is primarily regulated by the cell's internal redox state, implying it lacks multicellular inputs. However, the formation of fimbriae, rigid extracellular appendages mediating multicellular and cell-surface interactions, induces an internal redox shift that fully represses \textit{flu} via \textit{OxyR} \cite{RN156}. This repression, linked to bulk production of sulfhydryl groups in fimbriae, appears particular to fimbriation and is not observed with other extracellular appendages \cite{RN156}. In clonal communities, the internal redox shift caused by fimbriation is then a strong predictor of fimbriation in neighboring cells and the existence of fimbrial multicellular interactions. In effect, \textit{flu} expression is repressed by an internal change corresponding to a specific external multicellular interaction. Evolutionarily, \textit{E. coli} may have learned this particular redox shift was a useful way to sense fimbriation. Interestingly, regulation of \textit{flu} expression by reduced \textit{OxyR} is unaffected by hydrogen peroxide levels that induce the better-known operon regulated by the oxidized form of \textit{OxyR} \cite{RN149}. This illustrates a critical point about gene regulation: regulatory genes do not always have one-to-one mappings to inputs and can control different outputs depending on differences in input signals. Identifying one inducing condition for a gene network does not strictly rule out others. Fimbriation is first detected around the ~8-cell stage of self-organization \cite{RN36}, indicating \textit{flu} repression would occur by this point. While fimbriae extend rigidly to 2 $\mu$m, \textit{Ag43} binds cognates only within approximately 10 nm. Consistently, fimbriae can completely block \textit{Ag43} interactions \cite{RN80, RN156}, suggesting little benefit to separately repressing \textit{flu} expression. Polymerization of fimbriae takes time, and they are initially absent at the new cell poles formed by division: precisely where \textit{Ag43} mediates sister-cell adhesion \cite{RN213}. \textit{flu} repression by fimbriation may targetedly reduce polar adhesion between sister cells, perhaps to promote detachment of multicellular clusters and initiate new rounds of multicellular chain formation. Such detachment and chain initiation was observed directly and may facilitate self-propagation of \textit{E. coli} biofilms across surfaces via chain morphogenesis \cite{RN36}.

Cluster detachment does not robustly occur in every instance of self-organization however, suggesting there may be other reasons for \textit{flu} repression. During rosette formation, self-propelled motion causes adhered sister cells to reposition relative to each other \cite{RN37}. Adhesion by \textit{Ag43} between the outer membranes of sister cells would then produce mechanical tension on both cell envelopes. This tension would substantially increase about 10 minutes after the first cell division, coinciding with the onset of flagella-driven cell motion \cite{RN37, RN39}. The mechanical forces on individual bacterial cells have been reviewed \cite{RN157}, but tensile forces between moving sister cells have received little attention. Such tension might reflect environments that permit adhered cell motion, and would be absent in turbulent settings that immediately separate sister cells or in forced organization where cells experience compressive forces \cite{RN39}. Hence, tensile interactions between sister cells might serve as a behavioral sensor for the mechanical properties of their surroundings.

Intriguingly, similar mechanisms play developmental roles in higher organisms \cite{RN166}. Cadherins \cite{RN162} are functional analogs of \textit{Ag43} in metazoa, but evolutionarily unrelated: they are self-recognizing extracellular proteins that bind their cognates on neighboring sister cells. These and related proteins are thought to have played a key role in the evolution of multicellularity \cite{RN159, RN158, RN160}. During development in higher organisms, mechanical tension between adhered cells is transmitted via cadherins to cytoskeletal rearrangements that determine cytokinesis orientation \cite{RN161}. From a multicellular-graphs perspective, cadherin-mediated tensile interactions (edges) define how and where new cells (nodes) are added during graph propagation.  \textit{E. coli}'s cytoskeleton is not as flexible, and possible sensing mechanisms must differ. \textit{E. coli} and other gram-negative bacteria have two plasma membranes separated by a periplasmic space tens of nanometers wide (see \cite{RN163}). Tension between cells whose outer membranes remain in contact should dynamically expand the periplasm.  Osmotic shifts can cause similar fluctuations and several two-component systems sense such periplasmic changes. For example, the \textit{Rcs} system \cite{RN164, RN165} plays a critical role in biofilm formation \cite{RN88}, regulates many \textit{E. coli} multicellular genes, and can be activated by hydrostatic or osmotic conditions. Recent insights suggest that an inner membrane component of this signaling cascade is blocked by an outer membrane protein and signaling is triggered when the two membranes are pulled apart \cite{RN165}: the expected consequence of \textit{Ag43} interactions between moving sister cells. This is just one hypothesis for sensing mechanical cell-cell interactions and many ought to be explored, including other two-component systems which may respond to multicellular signals in addition to their identified envelope-disrupting signals. Hence, \textit{flu} expression is regulated by and may also generate multicellular interactions specific to self-organization. Its repression by fimbriation would then serve to down-regulate expression cascades triggered at the initiation of self-organization. Similar development-like regulatory feedback loops can exist entirely at the molecular level within the cell (see for example \cite{RN28}). The above considerations suggest \textit{E. coli} may incorporate external multicellular interactions into such feedback loops as well.

Following cell division, sister cells fold onto one another driven by the propulsion of newly synthesized flagella \cite{RN37, RN39}, the tail-like structures responsible for cell motility. A similar motion at the 4-cell stage completes rosette formation. Newly synthesized flagella exert lower forces than mature ones \cite{RN87, RN167, RN168} and can promote folding rather than separation which can result from stronger forces that overcome the adhesive interactions between cells \cite{RN37}. Expression of flagella biosynthesis genes must be tightly regulated for self-organization to occur: premature expression causes sister cells to separate, while delayed expression prevents rosette formation at the 4-cell stage \cite{RN37}. There are approximately 50 genes involved in flagella biosynthesis which are distributed across 17 operons \cite{RN172} and expressed in an ordered sequence, nicely illustrated in \textit{E. coli} using fluorescent transcriptional reporters \cite{RN42}. The flagella regulatory cascade is among the most extensively studied in bacteria and is induced on hydrostatic surfaces, partly explaining classic motility assays \cite{RN170, RN171}. Despite this detailed knowledge, the cellular or molecular signals triggering induction in \textit{E. coli} K-12 remain unclear. This has in part resulted from many motility findings being made with a derivative strain wherein the \textit{flhDC} promoter has been mutated to be induced by the absence of glucose \cite{RN169}, (providing a very useful experimental tool). Single-cell studies of the wild-type strain in "mother machine" devices \cite{RN173} suggest flagella gene induction occurs stochastically \cite{RN32, RN98}. As mentioned above however, this gene network is robustly induced during multicellular self-organization: at the two-cell stage, nearly all sister-cell pairs exhibit flagella-driven folding with similar dynamics \cite{RN37, RN39}.

Flagella gene expression requires cell division \cite{RN175} and is regulated in coordination with the cell division cycle \cite{RN87}. \textit{FlhDC} complexes induce the flagella gene expression cascade. In the absence of \textit{FlhC}, \textit{FlhD} disrupts the cell cycle \cite{RN86}, suggesting molecular interaction with \textit{E. coli}'s cytoskeletal components. During multicellular self-organization, flagella-driven motion begins approximately 10 minutes after the first cell division and is generally observed in only one cell of a sister-cell pair \cite{RN37}. This asymmetry is more pronounced in mutant strains locked into tumbling behavior, where one cell actively separates and tumbles away while the other remains immobile. Thus, \textit{E. coli} exhibits asymmetric flagella biosynthesis within sister-cell pairs during multicellular self-organization. This is reminiscent of asymmetric flagella synthesis in \textit{Caulobacter crescentus} \cite{RN174}, though methodological limitations had obscured possible similarities in \textit{de novo} flagella biosynthesis in \textit{E. coli} \cite{RN174}. Asymmetric division in bacteria is often considered a prototypical form of cellular differentiation \cite{RN176}, though this term can have more complicated connotations in higher organisms. From an evolutionary perspective, asymmetric division can enable division of labor, where cells perform distinct functions that benefit the overall community \cite{RN176}. From a graph-theory standpoint, asymmetric division produces two nodes (cells) with different behaviors that stably interact to reposition themselves and their descendants. Depending on one's theoretical lens, flagella gene expression during multicellular self-organization could represent cellular differentiation, division of labor, or a way to reliably propagate a dynamic multicellular graph. Asymmetric flagellation necessarily entails reduced sister-cell correlation, which may fit with the previous observations \cite{RN32}. The mechanism underlying this asymmetry remains unclear, but a clue emerges from the stochastic study: deletion of \textit{ydiV}, which binds \textit{FlhDC} complexes and inhibits their interaction with downstream promoters \cite{RN177}, increases correlation in flagella gene expression of sister cells, suggesting it is responsible for the asymmetry between them in wild-type cells. Environmental factors may also play a role. As noted, adhered sister cells may facilitate particular mechanical signals in hydrostatic environments, which are lacking in constrained conditions. The conditions in "mother machine" devices constrain cells but not completely and still allow some motion. Adjusting the physical parameters of these devices could tune flagella gene expression and ultimately help identify the precise signals inducing this gene network.

Although the exact signals inducing the flagella gene network remain unclear, its robust expression during self-organization profoundly shapes multicellular interaction networks. The arrangement cells in rosettes could enable both compressive and tensile forces between them. Due to the symmetry of \textit{E. coli} rosettes, each cell would be expected to experience equivalent interaction signals, potentially coordinating future gene network induction via mechanical sensing. Beyond physical contacts, rosette formation  uniquely facilitates multicellular interactions through diffusible molecules. It creates a tube-like inner cavity between cells where molecular concentration gradients would differ from the external environment. The significance of controlling such inner volumes and their diameters for communication by diffusible molecules was previously discussed \cite{RN39}, and the role of rosette formation in downstream multicellular gene expression is demonstrated by \textit{rpoS} transcription at the end of self-organization \cite{RN37}.  The self-organizing behavior resulting from robust flagella gene expression is fundamental to establishing rich and unique multicellular interaction networks. As indicated by mutants \cite{RN37} and analysis \cite{RN39}, such interaction networks cannot be reliably formed without properly timed flagella gene expression. Overlapping ideas exist regarding the necessity of multicellular rosettes to both multicellular development and evolution (discussed in \cite{RN37}). A more formal statement inclusive of these ideas is that rosettes are fundamental geometries for enabling chemical and physical interaction networks in a way that can be controlled as cells divide and communities grow (see also \cite{RN39}). 

From the 4-cell stage until there are approximately 1,000 cells when cell division ceases, \textit{E. coli} communities extend longitudinally as chains, preserving a constant width and elongating the inner tube-like cavity established by rosette formation. This extension depends on the induction of two key adhesins: type-1 fimbriae and curli \cite{RN36}. These structures are essential for biofilm formation but are typically repressed in single cells. Fimbriae are rigid protein appendages that extend 1–2 $\mu$m radially from the cell envelope \cite{RN179}. Their regulation is very complex \cite{RN178}, but is primarily determined by genomic inversion of the \textit{fim} operon's promoter between on and off states. Two recombinases mediate this switching: \textit{FimE} catalyzes the on-to-off transition, while \textit{FimB} catalyzes both directions \cite{RN83, RN180}. Molecular regulation suggests that numerous chemical and mechanical signals can influence expression of these recombinases, and elucidating the mechanism of fimbriation between the 4- and 8-cell stages will likely be challenging. Some possibilities stand out though. For example, \textit{Rcs} represses \textit{fimE} thereby promoting fimbriation \cite{RN181}, potentially in response to mechanical cues arising at the 2- to 4-cell stages (discussed above). Additionally, \textit{QseBC} induces the \textit{fim} operon \cite{RN178} and responds to the bacterial communication molecule autoinducer-3 (AI-3) \cite{RN183,RN182}. The internal cavity formed by rosettes produces barriers to diffusion that may unavoidably concentrate exported molecules like AI-3, thereby activating gene networks that single cells only induce in highly dense cultures (see discussion in \cite{RN39}). These considerations suggest fimbriation may be induced during self-organization by a combination of physical (\textit{e.g.}, \textit{Rcs}) and chemical (\textit{e.g.}, \textit{Qse}) multicellular interactions generated by the preceding multicellular stage. Engineering approaches in \textit{E. coli} have shown synthetic gene networks that are combinatorial regulated by two independent environmental inputs can produce robust digital-like expression \cite{RN184}. Perhaps \textit{E. coli} has already learned the same lesson and robustly expresses the \textit{fim} operon in response to orthogonal mechanical and chemical input signals that only co-occur within the multicellular interaction networks of self-organization.

Curli are extracellular protein fibers that polymerize from exported subunits \cite{RN185}, and their regulation is notably complex \cite{RN186, RN187}. In fact, the promoter of the curli master regulator \textit{CsgD} is considered to have the most complicated regulation among \textit{E. coli} promoters \cite{RN189}, with over 10 transcription factors influencing its expression.  Moreover, how its expression can result from the combinatorial processing of independent input signals (like the idea for fimbriae forward above) has already been illustrated \cite{RN189}. During self-organization, curli maintain the initial cell-cell configurations established by rosette formation and preserve the internal cavity. Curli’s role is first apparent at the 16-32 cell stage of self-organization and becomes more pronounced as community extension continue \cite{RN36}. In knockout mutants, the inner cavity is completely lost as is surface binding and the end of self-prorogation, even though cells remain adhered to one another. Further experiments are required to define the multicellular interaction network that regulates curli biogenesis during self-organization. However, two key examples may be particularly relevant. Curli expression is induced by low osmolarity through the \textit{EnvZ}/\textit{OmpR} system \cite{RN186, RN190}. Just as the concentration of excreted molecules could increase in the inner cavity of communities, imported molecules would become depleted \cite{RN39}, which depending on the molecule and its transport dynamics, could lower the internal osmolarity despite the external environment remaining unchanged. Separately, nutrient limitation influences curli regulation, notably through the master regulator of stationary phase genes, \textit{rpoS} \cite{RN186, RN191, RN192, RN193}. During self-organization, curli expression is detected many generations before \textit{rpoS} transcription becomes detectable, which occurs only just before attachment \cite{RN39}. This suggests curli may serve differing functions corresponding to different community stages. During multicellular extension, curli controls the diffusion properties within the multicellular community, similar to the role of the extracellular matrix in higher organisms.  At the time of multicellular dormancy, curli biogenesis may be further stimulated to combine with polyglucosamine (discussed below) and stably attach the cells in communities to each other and to surfaces.

Though the multicellular interactions that induce them remain imprecisely defined, fimbriae and curli expression fundamentally shapes subsequent interaction networks during multicellular self-organization. Through direct contact with neighboring cells, fimbriae may produce both compressive and tensile forces on the cell envelope. Additionally, as fimbriae extend to their mature length of 2 $\mu$m, they necessarily enlarge the inner cavity to a defined diameter, potentially enhancing molecular transport within this space. Curli meanwhile, which evidently  polymerize in the intercellular space, would necessary affect molecular diffusion in the inner cavity \cite{RN39}. Multicellular communication via diffusible molecules is essential for development across all organisms \cite{RN194}. Enabling such communication requires controlling extracellular concentrations of diverse molecules, which can be efficiently achieved by regulating the diameter of an inner cavity whose diffusion properties are adjusted by excreted polymers \cite{RN39}. During developmental processes in mammals for example, lumens (tube-like inner cavities) extend while filling with ECM polymers that selectively limit molecular diffusion \cite{RN195}. The biophysical principles underlying selective diffusion through ECM are common across organisms, including mammals and bacteria \cite{RN196}.  In \textit{E. coli}, constant-diameter extension of an ECM-filled inner cavity is therefore  expected to play a similar role in multicellular communication by diffusible molecules. This suggests fimbriae and curli expression enable multicellular interactions by the rich language of chemical communication throughout the subsequent stages of self-organization \cite{RN39}. The significance of this is demonstrated at the end of multicellular extension by the expression of \textit{rpoS} and \textit{pga} \cite{RN36, RN37}, both of which cause major phenotypic shifts and typically require substantial and independent chemical changes in the environment for their induction. Several division cycles occur between the initial evidence of curli biogenesis and \textit{rpoS} expression, yet the genetics of this interval remain unexplored. Given the dense behavioral program of the first several division cycles, many discoveries likely remain to be made during multicellular extension. Several possibilities have been proposed \cite{RN39}. For instance, bacterial communication via autoinducer-2 (AI-2) is a strong candidate \cite{RN39}, especially as it has been observed in microscopic communities of similar size \cite{RN188}. Another possibility is mating by conjugation between separate clonal multicellular communities \cite{RN39}. Given the establishment of an inner cavity for communication, the new multicellular interactions generated during multicellular extension are expected to primarily be chemical rather than mechanical. Discovering them will hence require transcriptional reporters to determine the expression dynamics of key genes that contribute to biofilms (\textit{e.g.}, \cite{RN68}) but whose roles in multicellular self-organization \cite{RN39} have yet to be identified.

\textit{rpoS} encodes an alternative sigma factor that regulates numerous genes involved in biofilm formation and cellular dormancy \cite{RN207}. It is the primary controller of gene expression during stationary phase \cite{RN205}, the portion of \textit{E. coli}'s life cycle where cells stop dividing and become senescent \cite{RN203, RN202}. \textit{rpoS} transcription is strongly induced shortly before stationary phase \cite{RN206}, though its expression is also subject to complex post-transcriptional regulation \cite{RN204, RN197}. During self-organization, \textit{rpoS} transcription, measured by a fluorescent reporter \cite{RN69}, is induced around the 200-cell stage.  This occurs approximately 60 minutes before communities attach to surfaces and reaches a relative maximum about 30 minutes before attachment \cite{RN37}. Under identical environmental conditions, including continuous delivery of fresh nutrients, \textit{rpoS} transcription did not occur in communities that were prevented from forming rosettes and the multicellular interactions characteristic of self-organization \cite{RN37}. \textit{rpoS} transcription during self-organization more strongly depends on the related multicellular interactions rather than the external environment. In addition to inducing transcription, self-organization may bring about key post-transcriptional inputs that regulate \textit{rpoS} expression, including hyperosmolarity and acidification \cite{RN204}, by propagating an inner cavity and controlling its diffusional properties. \textit{rpoS} expression is perhaps the most critical behavioral decision of \textit{E. coli}'s life cycle, and it is intriguing that a unicellular organism can use multicellular interactions to drive such a decision. Many of the genes expressed in stationary phase protect \textit{E. coli}'s internal molecular components (\textit{e.g.}, DNA, proteins, the periplasm, \textit{etc.}) \cite{RN203, RN202} without requiring sensing or adaptively responding to stresses. Protecting components in a way that does not require sensing and responding to external signals is a useful strategy for a community preparing to enter a terminal state \cite{RN39}. In a sense, \textit{rpoS} expression would therefore allow cells to ignore future external signals and insulate its gene networks from possible inputs. Moreover, \textit{rpoS} also down-regulates genes for fimbriation \cite{RN212} and flagellation \cite{RN211}, indicating negative feedback loops on the preceding multicellular interactions of self-organization.

\textit{rpoS} transcription, from the same fluorescent reporter, exhibited noisy behavior in "Mother machine" devices, corresponding to noisy cell growth \cite{RN35}. During self-organization, robust,  switch-like induction was observed without signs of heterogeneity or noise \cite{RN37}. These differences likely arise from how these devices produce input signals for the \textit{rpoS} network. As noted above, stochastic transcription in the "Mother machine" from the inducible \textit{rpoS} promoter may indicate its inducing conditions are incompletely met. Given the geometric constraints it imposes on cells, the "Mother machine" \cite{RN173} would not be expected to optimally induce genes regulated by multicellular inputs, resulting in noisy expression. This may indicate a larger issue however. "Mother machine" and similar micro-channel devices have been widely used for single-cell bacterial studies, hundreds of which report examples of either gene expression or cell behavior as noisy or stochastic. Increasingly, more and more cell behavioral decisions appear to be random or unpredictable. A degree of noise is attributable to the molecular nature of gene expression \cite{RN133}, but bacteria have spent billions of years evolving the specific gene networks underlying their behavior. Perhaps at times, the noisy behavior observed in these devices  reflects their unanticipated affects on input signals for inducible gene networks, rather than inherent properties of the networks themselves.

Multicellular communities attach to surfaces and stop growing 30 minutes after maximal \textit{rpoS} transcription is observed \cite{RN37}, concluding self-organization in attached dormancy \cite{RN39}. Attachment depends on the synthesis and external export of poly-$\beta$-1,6-N-acetyl-D-glucosamine (PGA) \cite{RN36}. PGA is essential for cell attachment in biofilms \cite{RN200}, and its enzymatic hydrolysis disassembles biofilms for a variety of species \cite{RN199}, suggesting a promising therapeutic strategy. The genes responsible for PGA biosynthesis and export are encoded by the \textit{pga} operon \cite{RN200}, which is post-transcriptionally regulated by \textit{csrA} (RNA-binding "carbon storage regulator") \cite{RN201}. \textit{csrA} also post-transcriptionally controls activators of \textit{pga} transcription \cite{RN209}. Attachment and dormancy coincide at the conclusion of self-organization, as demonstrated by superimposing the quantified growth and motion data for communities \cite{RN36}. However, motion decreases earlier and more gradually than growth rate \cite{RN36} which appears to correspond to the switch-like induction of \textit{rpoS} \cite{RN37}. This suggests that \textit{pga} induction may precede \textit{rpoS} transcription during self-organization, though its effect is more evident later, after sufficient PGA is produced to fully immobilize communities. This timing may also shed light on the trigger for \textit{rpoS} transcription, which remains unclear despite considerable molecular regulatory knowledge \cite{RN210}. It has long been known that \textit{rpoS} transcription is induced by reduced growth \cite{RN206}; the considerable metabolic cost of PGA synthesis \cite{RN200,RN198} may generate a sufficient inducing signal during self-organization. Coordination between \textit{rpoS} and \textit{csrA}, which regulates \textit{pga} genes, would be likely for another reason as well: \textit{csrA} represses several genes in the \textit{rpoS} regulon which must therefore be de-repressed in order to be expressed in stationary phase. Simultaneously tracking \textit{rpoS} and \textit{pga} expression in communities will be required to clarify the sequence of induction events during self-organizing.  PGA effectively freezes the relative position of cells in communities and thereby stabilizes the terminal stage of multicellular graph propagation \cite{RN39}, firmly restricting any changes in multicellular interaction networks.  Moreover, the extracellular polymer matrix PGA produces acts as a diffusional barrier that may serve to insulate \textit{E. coli}'s gene networks from external signals. Similar to the effects of \textit{rpoS} expression on input signals, PGA expression therefore helps multicellular communities tune out changes to their environment.

Multicellular interaction networks are inevitable in communities: cells generate chemical and physical interactions, intentionally or not. As \textit{E. coli} self-organization demonstrates, multicellular interaction networks can serve as both the inputs and the outputs of gene networks (\textbf{Fig. 4}) and they are dynamically rearranged by the behaviors of individual cells. Cell behavior can then predictably determine the inputs for gene networks, enabling self-organizing multicellular communities to control their own gene expression. This basic concept of \textit{multicellular control of gene networks}, where self-generated multicellular interactions produce regulatory inputs for gene networks, is a straightforward consequence of gene regulation \cite{RN1} within the dynamic-graph framework for multicellularity \cite{RN39}. It can therefore serve as a general model for predictable gene expression in communities and for sophisticated processes like cell differentiation and multicellular development. Multicellular interaction networks are not static: they change during self-organization and hence must be identified by directly observing the organizational behaviors of individual cells rather than extrapolated from single-cell properties. That is to say, the dynamic multicellular interaction networks matter, not just the static gene networks. Even in \textit{E. coli}, documenting, analyzing, and simulating such networks will be required to reveal how gene networks and cell behavior are connected.  While gene regulation helped molecular biology explain cell behavior, \textit{multicellular control of gene networks} indicates the reverse idea in self-organizing communities: cell behavior helps explain the complicated networks of molecular interactions inside each cell.

\begin{figure}
\includegraphics{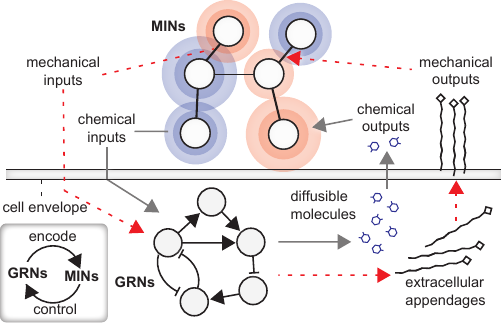}
\caption{Multicellular interaction networks (MINs) can serve as both inputs and outputs for gene regulatory networks (GRNs).  Multicellular interaction networks dynamically control gene networks and are also encoded by them.}
\end{figure}

Like many current microbes, \textit{E. coli} and its microscopic multicellular behaviors evolved in a world dominated by large multicellular organisms (\textit{e.g.} animals) that produce macroscopic forces. Some environments lack such forces (\textit{e.g.} inside host cells or on hydrated surfaces) and encourage multicellular growth by virtue of not forcibly separating cells. For many microbes, being in a multicellular community may better predict a particular environment than other external signals, \textit{i.e.} \textit{multicellularity} can represent an ecological niche as a whole. Useful genes for these environments (\textit{e.g.},  those encoding toxins, antibiotic tolerance, or other virulence factors) would naturally be under \textit{multicellular control}. This may partly explain the expression of such genes in macroscopic multicellular biofilms, even when large-scale communities are not part of pathogenesis \textit{per se}. In the future, infections might be prevented or treated by targeting the specific multicellular interactions that control expression of a pathogen's virulence factors. Mechanistically-targeting interactions will likely prove more feasible than interrupting community formation in its entirety. \textit{Multicellular control} may separately suggest how unicelluar organisms evolve simple multicellular behaviors: the multicellular interactions that are facilitated by an environment become connected through gene networks to regulate genes that improve fitness in that environment. Microbes could evolve multicellular behaviors by rearranging their gene networks to robustly control gene expression in given environments. As noted, applying dynamic graphs to multicellular self-organization indicates general constraints for community propagation, independent of the environment \cite{RN39}. Such constraints, in combination with environmental ones, are expected to guide the diverse multicellular behaviors performed by microbes. Beyond bacteria, such behaviors have also been investigated in fungi \cite{RN130, RN58}, protists \cite{RN119, RN121, RN132, RN120}, and even archaea \cite{RN122}.

Multicellular interactions can be functionally separate from the genes they control: in theory, any cell-cell interaction, chemical or physical, that is sensed can enable predictable gene expression and be adapted for \textit{multicellular control}. A gene’s multicellular role would differ from its unicellular role (\textit{e.g.} flagella propulsion in \textit{E. coli} self-organization versus single-cell motility) and must be directly observed. The adaptability of existing genes for multicellular interactions also suggests multicellular behavior can evolve by re-purposing genes with unicellular roles rather than requiring \textit{multicellularity genes}. Similar ideas have been explored in the evolution of animals \cite{RN107}. Multicellular control of gene networks could hence arise any time a species senses the physical and chemical interactions it already forms to then regulate genes, suggesting a model for step-by-step evolutionarily innovation of multicellular behaviors and complexity from existing genetics. Many genes would remain independent of \textit{multicellular control}: the \textit{lac} operon for example helps \textit{E. coli} consume lactose that mammalian infants receive, and might be best regulated solely based on the nutrients presence. Additionally, interaction- and gene networks that become adapted for \textit{multicellular control} could lose some of their original unicellular functionality (\textit{e.g.}, \textit{E. coli} K-12 is considered a poor swimmer due to its conservative regulation of flagella).  Creating a new multicellular behavior by rearranging a gene network could sacrifice an organism's unicellular fitness in a given context. Unicellularity has had its advantages for biological organisms, and this tradeoff may constrain how microbes introduce new multicellular behaviors into their life cycles. From this perspective, obligate multicellularity in higher organisms is a logical consequence following from their extensive use of multicellular control of gene networks; \textit{i.e.}, too many of their gene networks have become unsuitable for unicellular survival, except in specific conditions like tissue culture.

Multicellular communities are inherently dynamic: the number, positions, and interactions of cells constantly change. To be predictable, communities must constrain their own dynamics. From the dynamic-graph perspective, this means controlling the rules of propagation between multicellular graphs \cite{RN39}. \textit{Multicellular control of gene networks} provides a mechanism for accomplishing this and can be used to alter the propagation rules at one stage in anticipation of future stages, thereby ensuring simple, predictable dynamics as a communities grows. As suggested by \textit{E. coli} self-organization, multicellular interaction networks can be arranged such that the gene-network outputs from one stage define the gene-network inputs for later stages, creating a dynamic \textit{multicellular daisy chain}.  Daisy chains are engineering motifs where the outputs of one stage serve as the inputs of another (usually in a static rather than dynamic sense), and they have been applied to control the self-propagation of gene drives \cite{RN108}. \textit{Multicellular daisy chains} though are a class of dynamic graphs in which the propagation rules are determined by cells' gene expression responses to preceding organizational stages. Sensing an interaction network at one multicellular stage leads to a gene expression response that then creates a new interaction network to be sensed by a subsequent stages, and so on. As illustrated (\textbf{Fig. 5}), multicellular daisy chains are a simple control structure for reliable propagation of multicellular graphs during clonal self-organization: the gene-network inputs at each stage are predictably specified by the dynamically formed multicellular interactions of preceding stages. This enables a community to control gene expression with relative robustness (or indifference) to fluctuations in the external environment. Daisy chaining explains how otherwise noisy genes are arranged in an ordered sequence of expression stages during \textit{E. coli} self-organization. As pointed out, key behaviors (\textit{e.g.}, adhesion, repositioning, volume control, \textit{etc}.) are required for the reliable propagation of multicellular graphs \cite{RN39}. Daisy chaining by multicellular control of gene networks allows these requirements to be addressed in a hierarchical manner (as seen in \textit{E. coli} \cite{RN39}), supporting the idea that multicellular behaviors are "well-ordered" (\textit{i.e.} contingent on one another) \cite{RN39}. Daisy chains may also help disentangle \textit{E. coli}’s multicellular gene networks which are highly complicated when all regulatory interactions are superimposed and considered statically (see \textit{e.g.} \cite{RN129,RN128}).  During multicellular self-organization, the inputs for these networks change and are dynamic rather than static. Daisy-chain models would spread out the timing of these inputs, clarifying when and how they determine, and are determined by, cell behaviors.

\begin{figure}
\includegraphics{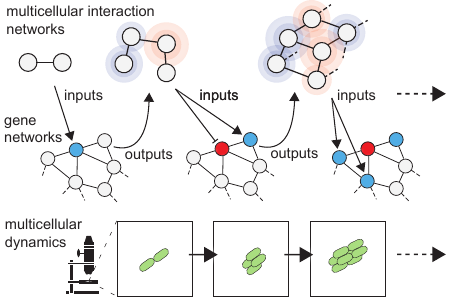}
\caption{Multicellular interaction networks allow gene-network outputs from one stage to specify inputs for later stages, thereby creating multicellular daisy chains.  Such daisy chains give biological organisms a mechanism to predictably control the propagation of their multicellular graphs and therefore the behavior of each of their cells during development. Deciphering the inputs and outputs of multicellular daisy chains may help simplify the complexity of biological organisms.}
\end{figure}

The idea that community structure drives gene expression and \textit{vice versa} is accepted for complex organisms, though how to best frame this idea is less clear. Dynamic daisy chains, composed of gene- and interaction networks, provide a novel and general framework for developmental processes (both empirically and theoretically). As they can be applied to any form of multicellular self-organization and can depict causal connections, daisy chains could prove very useful for understanding higher organisms. How to map developmental details to daisy chains may be immediately apparent to experts in many cases, and gene-network inputs and outputs have been thoroughly studied for development in many organisms. Reverse engineering daisy chains, \textit{i.e.} elucidating the stage-by-stage connections between their inputs and outputs, would help narrow the focus for investigations of multicellular organisms. An organism may employ many forms of multicellular interaction, but the robustness of developmental processes suggests that only a small set likely serve as inputs for any given stage, possibly reducing the number of cause-effect relationships needing to be deciphered. The identities of cells change during development, and the way they respond to given multicellular interactions would change as well: \textit{i.e.}, input-output connections are also stage-dependent. The daisy chain model for development may prove useful in the future for reverse engineering complex organisms and rendering their cell behaviors predictable.

Multicellularity is not arbitrarily complex, and the daisy chain model suggests clear limits. Each organism has a finite set of gene networks, thereby circumscribing the number of multicellular stages that can be arranged as a daisy chain. Hence, every daisy chain has a given set of regulatory stages, including a first and last. By definition, cell division past the last stage is uncontrolled and would produce unpredictable multicellular-interaction inputs for gene networks, leading to unstable multicellular graphs. Uncontrolled cellular growth, \textit{e.g.} tumorigenesis and cancer, is disastrous for multicellular communities as it undoes the ordered dependability that is the hallmark and culmination of their self-organization. An obvious solution is for a daisy chain's last stage to establish a terminal multicellular graph \cite{RN39} where cell division mostly stops and the existing multicellular interactions are stabilized.  This enables multicellular organization to be entirely controlled within the context of a daisy chain. Dividing cells are far from thermodynamic equilibrium, presenting a fundamental challenge to the dynamic stability of organisms; terminal multicellular graphs address this challenge by approximating an equilibrium state, \textit{i.e.} one with local dynamic stability. Terminal multicellular graphs are synonymous with terminally differentiated cells, which define the mature state and properties of complex multicellular organisms. Intriguingly, \textit{E. coli} also ends self-organization with a terminal graph where cells cease dividing and stabilize their multicellular interactions \cite{RN36,RN39}, though cell-cell heterogeneity seems unlikely given the uniform transcription of stationary-phase regulator \textit{rpoS} \cite{RN37}.

Terminal graphs may also have significance for the evolution of multicellular complexity; without a controlled conclusion, complex self-organization would have little value. For higher organisms, gene networks for terminal graphs would then have evolved before many of the developmental stages that precede them. Biological complexity would more likely arise from adding new intermediary stages or branch points prior to terminal graphs, rather than adding stages to the end of an existing daisy chain. As the gene networks and multicellular interactions that organize daisy chains are genetically encoded, an organism's concluding terminal graphs are predetermined before it begins multicellular self-organization. That is to say, a multicellular organism's characteristics, often viewed as emergent properties arising from systems-level interactions between cells, are at the same time the endpoints of programs which are specified in each individual cell. That single cells like zygotes reliably create entire organisms with common properties indicates that biology is highly predictable in some regards. Elucidating the daisy chains that lead to terminal graphs may bring this predictability more clearly into focus.

Multicellular daisy chains are modular and can be separated dynamically into individual stages. “Artificially” approximating multicellular interaction networks would be expected to induce multicellular genes even when natural processes of development or self-organization were not followed. For example, forcing multicellular organization by surface growth or aggregation could establish some of the interactions of clonal multicellular self-organization and would affect gene expression accordingly. This provides a surprising but plausible alternative explanation for \textit{E. coli}’s expression of multicellular genes when grown as colonies on agar \cite{RN131} or other surfaces even when cells have limited ability to self-organize. Precisely what \textit{E. coli} does in its primary habitats is an open question and the answer likely blends aspects of constrained- and self-organization \cite{RN36}. Potentially any multicellular community might induce expression of some genes under multicellular control, even outside an organism's native inducing environment.  The underlying logic is analogous to that of organoid models \cite{RN92}, where “artificial” multicellular communities that approximate multicellular interaction networks can induce cell-scale changes associated with organs, despite not resulting from natural organogenesis. In the future, further deciphering the sensible interactions in naturally- and artificially-organized multicellular communities will improve  biological control strategies like organoid models and “synthetic development" \cite{RN92, RN125, RN114, RN126, RN115, RN109,  RN124, RN144}. Correctly approximating the right intermediate stages of multicellular daisy chains might be key to creating tissues and organs in the lab. These examples also suggest how multicellular control of gene networks might explain sophisticated developmental processes in aggregative, rather than strictly clonal, multicellularity.  If cell densities are regularly high enough to facilitate aggregation, devolution might lead an organism to lose the clonality constraint for self-organization \cite{RN39} while keeping the downstream multicellular daisy chain stages encoded in its gene networks.

Cells in multicellular organisms often cooperate rather than cheat. Discourse on this point tends to conceive of cells as individuals capable of behaving selfishly to derive benefits at their community’s expense \cite{RN138,RN127, RN137}. Ensuring cooperation and deterring cheating is therefore thought to be a key challenge for multicellular organisms and their evolution. In the case of clonal multicellular self-organization however, where cheating is rare \cite{RN137}, a community’s entire behavioral sequence (\textit{i.e.} its daisy chains) and each cell’s identity is effectively predetermined. Cells in these communities are not individuals \textit{per se}; they are manifestations of the genetics and behaviors of the cells that came before them, going back to the lone individual that initiated multicellular self-organization. Daisy chains allow gene expression in each cell to be predictably control by multicellular interactions. The robustness of this expression is further enhanced by bistable genetic, epigenetic, and other regulatory switches incorporated as part of gene networks. In this context, cheating may not be an option: each cell and its cooperative behavior with others is already determined by the control logic of its multicellular daisy chain. As part of their life cycle, many organisms have a unicellular phase in order to initiate new multicellular communities; from the dynamic-graph perspective, it was suggested such a unicellular start is critical to the self-organization of complex organisms \cite{RN39}. Communities overrun with cheaters have limited control over their own members and propagation, limiting their potential for organizational sophistication.

To be general, principles of multicellularity must apply to all living organisms, from simple to complex. In part, the complexity of an organism depends on how multicellular interactions are formed, sensed, and controlled and the number of multicellular stages arranged as daisy chains. For consistency, multicellular simplicity, \textit{e.g.} unicellularity, must result from the lack of specific multicellular interactions or behaviors. Studying multicellular interactions mechanistically in higher organisms is difficult for a number of reasons. For example, to canalize developmental processes \cite{RN136, RN111}, higher organisms may often rearrange gene networks to be independent of the multicellular interactions that originally controlled them. Canalization increases the robustness of development but, in so doing, can mask the multicellular interactions that were necessary for evolution of complex organisms. Alternatively, microorganisms with relatively simple multicellularity can be helpful for uncovering general principles, and their lack of evolved sophistication means key multicellular interactions have not yet been hidden. \textit{E. coli}’s documented self-organizing behaviors suggest it could serve as a useful model organism for the investigation of multicellular biology and exploration of principles of multicellularity. Engineers often compose complex systems from simple motifs that individually control particular behaviors. Similarly, complex organisms use a collection of multicellular behavioral motifs which can also be seen in simple organisms (\textit{e.g.}, rosette self-organization). If multicellularity is well-ordered as speculated \cite{RN39}, these behavioral motifs can be studied one-by-one and the causal connections between them can be discovered experimentally. Problems in biological complexity might then be reduced to set of multicellular behavioral motifs. Re-wiring the gene-network inputs and outputs of such behavioral motifs could enable predictable composition of new multicellular behaviors and daisy chains, providing a theoretical basis for efforts in synthetic multicellularity \cite{RN112, RN113, RN54, RN115,RN116,RN117}. To an extent, turning synthetic biology into a predictable engineering-like discipline will first require elucidating the predictable engineering-like principles of multicellular biology. Starting small, with multicellular motifs, may be the best way to do this. Beyond explaining why some organisms are complex and others are not, a general theory of multicellularity would provide a framework for understanding biological organisms, their limits, and the causes of their diseases and dysfunctions.

Multicellular graphs and the living communities they represent are like crystals in motion. Their structures reorganize in time to alter cell interactions and identities. Framing multicellularity by dynamic graphs partly resembles quantum chemistry, where nodes (atoms) are connected by edges (bonds) that follow specific interaction rules (electron orbital theory). Cells of course complicate community structure by dividing: the nodes and edges of multicellular graphs continually change until ultimately reaching a terminal state. Rather than letting their communities become too complicated or unpredictable, cells simplify things by changing their interaction rules through gene regulation to anticipate their own futures. This allows them to reliably propagate multicellular graphs and in so doing predictably control the behavior of each cell, stage by stage. The blueprint for clonally self-organizing multicellular communities, both the gene networks and interaction networks that create daisy chains, is encoded in each cell’s DNA. But genetics alone are insufficient to understand multicellular self-organization which is driven by the changing interactions between cells and must be discovered by directly tracking the cells in communities. Unlike quantum chemistry, these nodes and the changing interactions between them can actually be observed with the help of optical microscopes. The necessary experiments can be challenging, and the effort and patience they require limit their allure. At least, due to modern machine-assisted techniques, we no longer need to press our eyes against microscope lenses for hours just to learn what bacteria like \textit{E. coli} do. I wonder what Andre Lwoff would think about all of this, though he may have already told us: "I decided to operate with individual bacteria...I like to see things, not calculate probabilities" \cite{RN72}. To understand how gene networks produce cell behavior, we must \textit{see} how cells self-organize their multicellular interactions. True for \textit{E. coli} and true for higher organisms, only more so.

Evolutionary explanations often help animate our knowledge of bacteria, even if they are not always true. Evolution itself is a history of particulars: what exists directly depends on what existed; yesterday's outputs serve as today's inputs. Dynamic graphs illuminate multicellular properties of biological organisms that both expand and constrain their evolutionary possibilities \cite{RN39}. Someday soon, a new theory of multicellularity might explain evolutionary biology rather than the other way around. Here's looking at you, \textit{E. coli}.

\vspace{8 mm}
\textbf{ACKNOWLEDGMENTS}: The author thanks Joanna Goldberg for comments on early figure drafts and Monica Farley, David Stephens, and Carlos del Rio for their support.

\bibliography{SBA}
\end{document}